# Efficiency, Expressivity, and Extensibility in a Close-to-Metal NPU Programming Interface

Erika Hunhoff‡†, Joseph Melber*, Kristof Denolf*, Andra Bisca*, Samuel Bayliss*,
Stephen Neuendorffer*, Jeff Fifield*, Jack Lo*, Pranathi Vasireddy*, Phil James-Roxby*, Eric Keller†

*AMD, †University of Colorado at Boulder, USA, ‡Work performed at AMD
{erika.hunhoff,eric.keller}@colorado.edu {joseph.melber,kristof.denolf}@amd.com

*Abstract*—Accelerators such as neural processing units (NPUs) deliver an enticing balance of performance and efficiency compared to general purpose compute architectures. However, effectively leveraging accelerator capabilities is not always simple: low-level programming toolkits may require substantial developer effort while high-level programming toolkits may abstract critical optimization features.

This work aims to increase efficiency of designers using IRON, a toolkit for close-to-metal NPU performance engineers. We provide an updated programmer interface to IRON containing new and refined programming constructs. The new interface includes extensible features for placement and data transformation. These contributions are evaluated in terms of 1) efficiency, with analysis showing ∼26% average reduction in lines of code and decreases in Halstead metrics for a variety of designs; 2) expressivity, demonstrating the new interface supports the wide range of features and patterns already supported by IRON; and 3) extensibility, illustrating the new tooling for placement and tiling can be extended to accommodate common use-cases.

## I. INTRODUCTION

Accelerators such as neural processing units (NPUs) can provide increased performance and higher efficiency compared to general purpose processors; however, utilizing accelerator capabilities effectively is not always simple [40]. One approach is to construct high-level frameworks comprised of operations and compilers that abstract details of the accelerator architecture. Examples of this approach for NPUs are AMD Ryzen™ AI Software [5], Intel® OpenVINO™ [34], and Qualcomm® Neural Processing SDK [38]. These platforms enable developers to build and deploy machine learning models trained in common frameworks such as PyTorch and TensorFlow. A second approach is to construct low-level toolkits and libraries that support creation of bespoke, optimized designs using accelerator-specific characteristics. An example of this approach is IRON, a close-to-metal, general purpose, open-source toolkit for performance engineers targeting AMD XDNA™ NPUs [28], [29], [31]. Each approach serves a different, but important, purpose.

In low-level toolkits, the programming interface contends with tension between providing tools for programmers to easily express their intent (designer efficiency) and exposing nuanced hardware capabilities required for clarity and precision (complexity). Although some tension is fundamental, specific trade-offs can be navigated through intentional interface design. This work details contributions to IRON that allow designers to represent designs plainly (*principle 1: efficiency*) without compromising designer ability to influence low-level decisions (*principle 2: expressivity*). These contributions also enable designers to create new tools to generate portions of designs (*principle 3: extensibility*). Central to our contributions is a new API implemented above the existing IRON programming interface. This API refines IRON programming constructs, provides a new interface supporting extensible tooling for placement (the matching of design constructs to physical resources), and includes a new supplemental data transformation library, also designed to support custom extensions. The new API coexists with existing IRON APIs, and is integrated into IRON as a core programming interface in the open source repository, https://github.com/Xilinx/mlir-aie.

Contributions to IRON are evaluated on the three principles of efficiency, expressivity, and extensibility using a set of 27 IRON designs, ranging from minimal (a data passthrough) to complex (streaming vision pipeline for edge detection). Demonstrating expressivity, it was possible to express all evaluated designs using the new API, with consistent performance between new and previous design implementations. Indicating increased efficiency for designers, we find a ∼26% average reduction of single lines of code (SLOC) and average reduction across Halstead metrics [14] across designs. Illustrating extensibility, we craft a custom placement generator and a custom data transformation generator using the new API, and examine the utility of custom extensions. Twenty-four example IRON designs utilize the placement tool. Five of the designs, including a `GEMM` design with complex runtime data movement tiling patterns, use the data transformation generator.

The remainder of the paper is structured as follows. Section II provides an overview of NPUs. Section III summarizes IRON. Sections IV-VII present the design, implementation, and evaluation of our contributions. Section VIII discusses related work and Section IX contains concluding thoughts.

## II. ARCHITECTURAL OVERVIEW OF AMD XDNA™ NPUs

This section provides an overview of AMD XDNA™ NPU architecture (Section II-A) and implications of the architecture for programming and optimization (Section II-B).

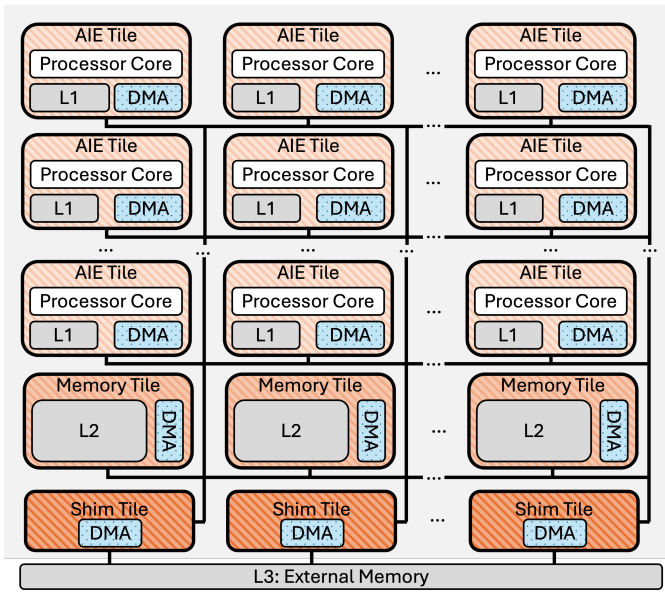

Fig. 1: Simplified diagram of an AMD XDNA™ NPU such as found in Ryzen™ 7040 processors [41].

### A. AMD XDNA™ NPU Architecture

An AMD XDNA™ neural processing unit (NPU) consists of several types of tiles arranged spatially in a two-dimensional grid connected by a streaming interconnect. NPUs are designed for power- and area-efficient inference: the AMD XDNA™ NPU found in Ryzen™ 7040 and 8040 SoCs provides >10 trillion operations per second (TOPS); using the XDNA™ programming stack for DNNs, models run with 4.3-33× performance per watt compared to the on-chip x86 processor [41]. The XDNA™ 2 architecture provides up to 50 TOPS [6].

Figure 1 is a simplified architectural diagram; comprehensive descriptions are available elsewhere [41]. AI Engine (AIE) tiles (also found in Versal™ architectures [1], [2]) contain a very long instruction word vector processor core with associated data memory, program memory, program counter, register file, and functional units [41]. Memory tiles provide additional on-chip SRAM scratchpad storage, while shim (or interface) tiles help manage external interfaces [41]. A streaming network-on-chip (NoC) supports point-to-point data transfers between tiles and can be configured for broadcast or selective multicast. Unlike architectures with hardware-managed caches, on NPUs on-chip data is buffered in scratchpads with software-managed data movement that is both explicit and deterministic [41]. Data movement accelerators (DMAs[1]) orchestrate asynchronous data movement, and using buffer descriptors (BDs), support reorder, reshape, and repeat patterns.

[1]DMA in this context refers to a physical component rather than the act of direct memory access.

### B. Implications for Programming and Optimization

While explicitly controlling data movement is an optional optimization possible in most cache-based architectures, the architecture of the NPU *necessitates* explicit data movement for all NPU designs. Explicit data movement can increase the determinism of application latency and throughput and can reduce power consumption [41].

The NPU architecture provides rich features to optimize data movement. Utilizing L2 buffers in memory tiles can reduce the off-chip bandwidth required for some applications. Duplicating data using broadcast supported by the NoC can further reduce off-chip memory bandwidth requirements. On-the-fly data transformations supported by DMAs can reduce the number of DMA operations needed to represent a data movement pattern. This is useful for partitioning data according to an algorithm and for structuring data appropriately for vectorized instructions. These attributes enable a performance engineer or compiler to fine-tune data movement.

The memory hierarchy and arrangement of AIEs in an NPU is spatial. In an NPU design, compute and data must be placed at explicit locations on the device. NPU designs are also temporal. Synchronization is supported in hardware by locks, and the integration of those locks with DMAs operations [41]. To allow a programmer to fine-tune how and when actions occur on the device, the programmer must be able to express both spatial and temporal relationships [27].

## III. IRON: A PERFORMANCE TOOLKIT FOR NPUS

This section summarizes IRON, an open-source programming toolkit for AMD XDNA™ NPUs [31]. IRON provides fine-grained control over spatial and temporal aspects of NPU designs, with an emphasis on data movement.

### A. IRON Overview

The heart of IRON is a multi-level intermediate representation (MLIR) dialect, `mlir-aie` [31]. MLIR is a technology that supports reusable, modular, and domain-specific compiler infrastructure [21], and is a popular choice for targeting diverse architectures, such as accelerators ([11], [44], [45], *etc.*). `mlir-aie` defines primitives and abstractions for NPUs with a focus on 1) clear and complete representation of hardware capabilities, and 2) simplified representation of critical design components such as data movement, synchronization, and design structure. `mlir-aie` is the core of the IRON low-level programming interface for NPUs and defines an IR that compilers for higher-level dialects may utilize to target NPUs.

Abstractions in IRON provide programming convenience by capturing common patterns but do not hide fundamental characteristics of NPU designs including placement, explicit data movement, resource limitations, and complex access patterns supported by DMAs. The abstractions in IRON are not targeted at a specific application domain, although some patterns supported by IRON are useful for digital signal processing and machine learning.

The core of the IRON Python API is auto-generated from `mlir-aie` via MLIR Python bindings [32]. Bespoke extensions to the bindings hide MLIR implementation details such as conversion between common Python types and MLIR types. IRON-specific extensions are supplemented by `mlir-python-extras`, an open source library providing helpers broadly useful for cleanly and succinctly representing MLIR in Python [26]. The IRON Python API supports metaprogramming by allowing a programmer to mix traditional Python with Python-driven MLIR operation generation.

### B. Lifecycle of an IRON Design

An IRON design can be written in Python using the IRON Python API or in MLIR using `mlir-aie`. If written in Python, the script will emit MLIR assembly. IRON designs use `core` blocks to specify the code to be executed on each processor core used by a design. Code in a `core` block may perform computations, call externally defined kernels, or both. External kernels are separately compiled using tools such as Peano [36]. `aiecc` is the primary compiler driver for IRON and takes as input MLIR assembly, along with any kernel object files, and produces two artifacts: a binary, which includes the programs to run as part of the design, and a sequence of instructions representing (re)configurations of the NPU to be loaded from a host processor during runtime. IRON interfaces with Xilinx® Runtime (XRT) to execute designs [47]. IRON designs are debugged using static analysis of generated MLIR and dynamic analysis through runtime tracing. IRON provides tools to configure tracing mechanisms supported by the hardware [3].

### C. Outline of an IRON Design

Shown in the example IRON design in Fig. 2a, all constructs that generate MLIR in an IRON design must be created within an MLIR `context` (line 1). At the end of a design, MLIR is emitted using the `module` constructed by the `context` (line 29). The first logical subsection of a design is a declaration of resources required (lines 4-9), including physical components (such as `tiles`) and IRON abstractions that are lowered to physical resources (such as `ObjectFifos`, which are described in Section III-D). Types are expressed using NumPy [15] types (lines 4-5).

`core` blocks are declared as decorated functions (line 11). The AIE acquires access to input and output buffers (lines 14-15) of tile dimension (`TH, TW`). Synchronization is controlled by the `ObjectFifos` with ownership of those buffers. The AIE performs element-wise scalar addition (lines 16-18) before releasing each buffer (lines 19-20). Loops are denoted with `range_`, a Python wrapper for a looping construct provided by the MLIR `scf` dialect [42].

The runtime sequence (lines 22-28) is used to send a sub-matrix (tile) of data, specified using sizes and strides calculated from the matrix and tile dimensions, to an `ObjectFifo` with a shim tile endpoint. The output tile is similarly received. The runtime sequence is a mix of declarations (`shim_dma_single_bd_task`), asynchronous operations (`dma_start_task`), and synchronous operations (`dma_await_task`).

### D. Data Movement in IRON

As outlined in Section II, a key architectural component of NPUs is explicit data movement. IRON supports data movement at varying levels of abstraction, but the primary abstraction is the `ObjectFifo`. An `ObjectFifo` is a circular buffer with a configurable amount (depth) of objects and has theoretical roots in dataflow theory [9]. Objects belonging to an `ObjectFifo` share shape and data type and are allocated at the discretion of the compiler. Accesses to `ObjectFifo` objects are mediated with locks managed by the `ObjectFifo`, and access is granted and revoked with `acquire` and `release` operations.

The `ObjectFifo` is designed to support common patterns such as broadcast, sliding windows (if n buffers are acquired, a call to `acquire(n + 1)` will acquire just one more; similarly, if m buffers are acquired, a call to `release(1)` will release just the first buffer, leaving the last m - 1 buffers acquired), pipeline balancing (the depth of `ObjectFifo`s in a critical path can be increased or decreased), splitting and joining data in L2 (using a `link` operation to join two `ObjectFifo`s over a shared memory tile endpoint), and L2 buffering (also using a `link`).

## IV. EFFICIENCY, EXPRESSIVITY, AND EXTENSIBILITY

Extensions to IRON were crafted in order to increase designer efficiency while maintaining designer expressivity and enabling extensibility.

*Efficiency* We seek to enable efficient expression of designs by allowing features of a design to be minimally expressed when that feature has not been selected for tuning by the designer. Designs are multifaceted, with many interesting aspects a performance engineer can choose to focus on (placement, dataflow, runtime operations, data tiling, kernels, *etc.*). Not all designs require a high level of detail in all aspects. Our contributions aim to allow a designer to efficiently express aspects of a design that are not the designers focus.

*Expressivity* We seek to gain efficiency – but not at the cost of expressivity. Our contributions to IRON should allow a programmer to express the same constructs as IRON, and designs written using new contributions to IRON should exhibit the same characteristics (including performance) as IRON designs expressed using existing IRON APIs.

*Extensibility* We seek to integrate pathways into IRON allowing performance engineers, who are not necessarily also compiler engineers, to generate portions of IRON designs using custom algorithms and tools. IRON provides full control over many aspects of a design, which makes it especially suited for use as an experimental sandbox for automation tools being developed for higher-level compiler workflows. Our contributions aim to enable integration of custom extensions for specific aspects of interest such as design space exploration, placement, tiling, and structuring of runtime operations.

```
1  with mlir_mod_ctx() as ctx:
2    @device(AIEDevice.npu1_1col)
3    def device_body():
4      mty = np.ndarray[(MH, MW), np.dtype[np.int32]]
5      tty = np.ndarray[(TH, TW), np.dtype[np.int32]]
6      shm0 = tile(0, 0)
7      aie2 = tile(0, 2)
8      fi = object_fifo("in", shm0, aie2, 2, tty)
9      fo = object_fifo("out", aie2, shm0, 2, tty)
10
11     @core(aie2)
12     def core_body():
13       for _ in range_(sys.maxsize):
14         a = fi.acquire(ObjectFifoPort.Consume, 1)
15         b = fo.acquire(ObjectFifoPort.Produce, 1)
16         for i in range_(TH):
17           for j in range_(TW):
18             b[i, j] = a[i, j] + 1
19         fi.release(ObjectFifoPort.Consume, 1)
20         fo.release(ObjectFifoPort.Produce, 1)
21
22     @runtime_sequence(mty, mty)
23     def sequence(dati, dato):
24       in_task = shim_dma_single_bd_task(fi, dati,
          ↪ sizes=[1, 1, TH, TW], strides=[0, 0, MW,
          ↪ 1])
25       out_task = shim_dma_single_bd_task(fo, dato,
          ↪ issue_token=True, sizes=[1, 1, TH, TW],
          ↪ strides=[0, 0, MW, 1])
26       dma_start_task(in_task, out_task)
27       dma_await_task(out_task)
28       dma_free_task(in_task)
29  print(ctx.module)
```

(a)

```
1  mty = np.ndarray[(MH, MW), np.dtype[np.int32]]
2  tty = np.ndarray[(TH, TW), np.dtype[np.int32]]
3  fi = ObjectFifo(tty)
4  fo = ObjectFifo(tty)
5
6  def core_fn(of_in, of_out):
7    a = of_in.acquire(1)
8    b = of_out.acquire(1)
9    for i in range_(TH):
10     for j in range_(TW):
11       b[i, j] = a[i, j] + 1
12   of_in.release(1)
13   of_out.release(1)
14 my_worker = Worker(core_fn, fn_args=[fi.cons(),
    ↪ fo.prod()])
15
16 rt = Runtime()
17 with rt.sequence(mty, mty) as (dati, dato):
18   rt.start(my_worker)
19   rt.fill(fi.prod(), dati, sizes=[1, 1, TH, TW],
      ↪ strides=[0, 0, MW, 1])
20   rt.drain(fo.cons(), dato, sizes=[1, 1, TH,
      ↪ TW], strides=[0, 0, MW, 1], wait=True)
21 print(Program(NPU1Col1(),
    ↪ rt).resolve_program(SequentialPlacer()))
```

(b)

Fig. 2: A matrix-scalar addition IRON design written without (2a) and with (2b) contributions.

## V. CONTRIBUTIONS TO IRON

Contributions to IRON are organized around a new API (Section V-1) which includes a refined `ObjectFifo` API (Section V-2) and new constructs for existing IRON design components (Section V-3). The new API supports an extensible placement interface (Section V-4) and includes a library for generating on-the-fly data transformations (Section V-5).

*1) Creating Distance from MLIR:* The structure of an IRON design is inherited from the structure required by MLIR. There are techniques to control and hide the `context` block (such as those provided by `mlir-python-extras` [26]), but if an MLIR operation is created when the corresponding Python object is created, then the Python objects are subject to the same rules as the corresponding MLIR operations. MLIR requires an operation be complete (*i.e.,* all important information known at construction), and be instantiated according to proper placement in the MLIR `context`. It is possible to insert and reorder MLIR operations in Python. However, we believe it would add complexity and make human comprehension of the IRON programming stack difficult if these techniques were used in both the compiler driver and the Python programming front-end, so we leave these techniques to `aiecc`.

The constraints of MLIR can be at odds with the goal of increasing the usability of IRON. For instance, one desirable improvement to IRON is reduction of duplicated information. An example of duplicated information in IRON designs are the placements of an `ObjectFifo` endpoint and the `core` block that uses the endpoint (Figure 2a lines 8-9, 11); the placement location could be provided once instead of twice. However, to construct the `core` block and `ObjectFifo`, both Python constructors must provide a placement at the time of instantiation – requiring information duplication.

To relax the requirements of position and information completeness, we create a new top-level IRON Python API. Python classes in the new API are inheritors of a `resolvable` interface and defer creation of MLIR operations until their `resolve` function is called. If the Python class does not contain enough information to resolve to an MLIR operation, `resolve` fails. Fig. 2b shows the design in Fig. 2a rewritten using the new API and associated contributions.

*2) `ObjectFifo` API:* Deferred resolution is used to refine the `ObjectFifo` API. `ObjectFifo` declarations (Fig. 2a lines 8-9) can be succinctly rewritten (Fig. 2b lines 3-4). The default depth of `ObjectFifos` is set to 2 (due to the prevalence of ping-pong buffers); the `ObjectFifo` name (which becomes the MLIR identifier) is auto-generated; and the `ObjectFifo` endpoints are inferred.

When an `ObjectFifo` is given to an operation or construct for use, an `ObjectFifoHandle` is generated by calling `prod` for a producer handle or `cons` for a consumer handle (Fig. 2b line 14). `prod` always return the same object, as an `ObjectFifo` can only have one producer. Multiple calls to `cons` generate multiple consumer handles, expressing a broadcast. Since a handle knows if it is a producer or consumer, there is no need to specify the `ObjectFifoPort` in `acquire` and `release`, further reducing redundancy (Fig. 2a lines 14-15, 19-20 vs. Fig. 2b lines 7-8, 12-13).

Instead of porting `link` to the new API, the `ObjectFifoHandle` class provides `forward`, `split`, and `join` methods which allow data to be forwarded over a tile (useful for L2 buffering), split across a tile (useful for splitting data staged in L2 in one L2-L3 flow to conserve channels), or joined across a tile (same use, but in reverse). At resolution, these methods ensure the generation of necessary `ObjectFifo`s and `link`s.

*3) Worker, Runtime, and Program Constructs:* We introduce a new construct, the `Worker` (Fig. 2b line 14). Construction of a `Worker` mirrors that of a thread in many multithreading libraries. A `Worker` takes a routine to run (`core_fn`) and the context (`fn_args`) needed to run it. Familiarity is one motivation for `Worker`s; a second motivation is the explicit separation of task definition (*i.e.*, the `core_fn`, Fig. 2b line 6) from the configuration for running the task (*i.e.*, arguments to a `Worker`, Fig. 2b line 14). This pattern provides a clear path for metaprogramming as `fn_args` can contain arbitrary Python values to customize the MLIR generated by a `core_fn`. This type of metaprogramming is often used to adapt computing tasks to variable datatypes, dimensions, or distributions of data.

A new `Runtime` construct represents the runtime sequence (Fig. 2b lines 16-20). A runtime sequence does not have the freedom to represent arbitrary computations, which can be an unclear distinction to IRON programmers as it is visually similar to the imperative block executed by AIEs in IRON designs. Operations supported by a `Runtime` include `start` to indicate a worker should be configured for execution (Fig. 2b line 18), `fill` to use a DMA operation to populate an `ObjectFifo` with data from a L3 buffer (Fig. 2b line 19), and `drain` to collect data from an `ObjectFifo` for a L3 buffer (Fig. 2b line 20). A last method, `inline_ops`, allows expert designers to insert a Python function generating arbitrary MLIR ops into the runtime sequence; this is roughly analogous to inline assembly in a C program. This technique supports bespoke configuration, and will be used in future work to fully support tracing configurations in the new API.

To compose these components into a design, the new `Program` construct creates a design from a `Runtime` applied to a `Device`. MLIR for the design is produced by `resolve_program` (Fig. 2b line 21).

*4) Interface for Placement:* The new API supports manually placed programs, meeting the goal of expressivity. However, if a programmer does not provide enough placement information, `resolve_program` will fail. Without additional extension, this requires all IRON designs be fully placed. To meet the goals of both efficiency and extensibility, we construct an interface supporting placement generation.

Objects in the new IRON API that are `Placeable` provide a `place` method that takes a placement tile as an argument. `resolve_program` optionally takes a `Placer` argument (Fig. 2b line 21). A `Placer` must implement `make_placement`. During the execution of `resolve_program`, `make_placement` is called, allowing the `Placer` to invoke `place` on all `Placeable` design components before MLIR is generated. The placement interface supports partially placed designs as well as placement hints `AnyShim`, `AnyMemTile`, and `AnyComputeTile`. Future work includes support for tagging `Placeable` components with programmer-defined attributes as a method of supporting arbitrary placement hints.

The interface for placement fulfills the goals of efficiency (designers uninterested in manual placement may use and reuse placement generators), expressivity (changing `Placers` or specifying partial placement for a given design is straightforward), and extensibility (a user has a well-defined mechanism to construct new `Placers`). The `SequentialPlacer` in Fig. 2b line 21 is an example `Placer` and is discussed further in Section VII-C1. The `Placer` API is written in Python but common language-binding tools can be used to wrap existing placement and design space exploration tools written in other languages into `Placers`.

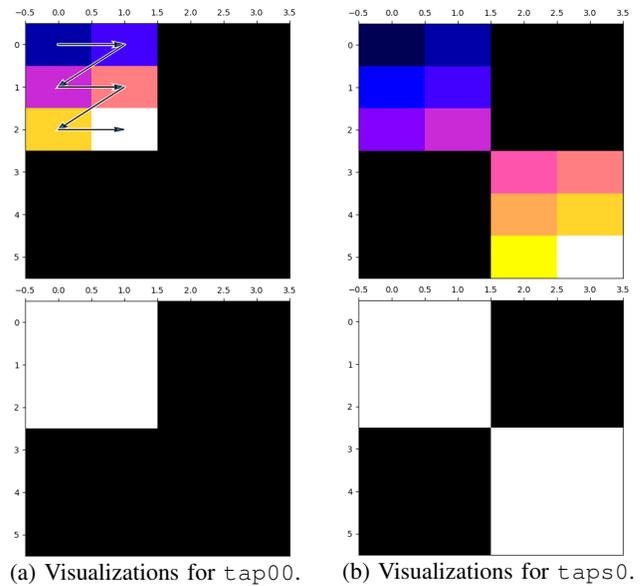

(a) Visualizations for `tap00`.  (b) Visualizations for `taps0`.

Fig. 3: Heat maps generated by `taplib` representing access order (top) and access count (bottom).

*5) Primitives for on-the-fly Data Transformations:* One point of complexity stands out in Fig. 2b: the sizes and strides (lines 19-20). Reasoning about transformations from just looking at sizes and strides can be difficult. To address this, we create `taplib`, a library containing two primitive abstractions for expressing and reasoning about transformations. A `TensorAccessPattern` (`tap`) is constructed from a set of tensor dimensions, sizes, strides, and an offset into the tensor. A `TensorAccessSequence` (`tas`) is a collection of `tap`s whose tensor dimensions match. Consider an example where a programmer desires to tile a $6 \times 4$ tensor into four

$3 \times 2$ non-overlapping row-major tiles and select the upper-left and lower-right. Using `taplib`, this can be expressed as:

```
tap00 = TensorAccessPattern((6, 4), offset=0,
↪ sizes=[1, 1, 3, 2], strides=[0, 0, 4, 1])
tap11 = TensorAccessPattern((6, 4), offset=14,
↪ sizes=[1, 1, 3, 2], strides=[0, 0, 4, 1])
taps0 = TensorAccessSequence.from_taps([tap00,
↪ tap11])
```

`taplib` provides mechanisms to reason about `ttaps` and `tases`, including visualizations and animations for 1- and 2-dimensional tensors (Figure 3). `taplib` also supports tools for programmatic analysis. The graphics shown in Figure 3 were produced using `taplib` access maps. An access count map is a tensor that contains, in each element, the number of times that a `tap` or `tas` accesses the corresponding element in the original tensor. An access order map is a tensor that contains, in each element, the element-wise enumeration of the order of accesses defined by the `tap` or `tas`. Access maps are useful for catching bugs, as a logical error in `tap` definition can be formalized as a `tap` whose access pattern, in count or order, does not match the intention of the programmer. The following code snippets demonstrates verification of `tap00` and `taps0` characteristics using access maps:

```
# Num accessed: Count starts at 0, so highest is 5
assert tap00.access_order().max() == 3*2 - 1
# Count of elements accessed by tap00
assert tap00.access_count().sum() == 3*2
# Num accessed: Highest count for two tiles is 11
assert taps0.access_order().max() == 2*(3*2) - 1
# The tas does not access any element more than once
assert taps0.access_count().max() == 1
```

We define a new concept of equivalence for access patterns using access maps. The strict definition of equivalence requires the tensor dimensions, sizes, strides, and offset be equivalent between two `taps`. We use the term *access equivalent* to describe `taps` which generate identical access order and count maps. Access equivalence is useful in the context of IRON because DMA units within an NPU have differing constraints on the maximum dimension of sizes and strides supported, as well as the maximum size (in bits) supported in each dimension. Thus, two access patterns may be strictly unequal, but be access equivalent, where one yields a valid IRON design and the other does not. `taplib` allows a programmer to express these intricacies precisely. The new IRON API integrates `taplib` (as an example, the Runtime `fill` and `drain` methods take a `tap` in lieu of sizes and strides).

## VI. IMPLEMENTATION

Our contributions are implemented in Python, with the new top-level API implemented in ~1,400 lines of code (LOC), and `taplib` implemented in an additional ~560 LOC. The effectiveness of the contributions are largely agnostic to the implementation mechanism; some contributions are suitable for integration into `mlir-aie` as a matter of future work.

## VII. EVALUATION

Our contributions aim to increase designer efficiency (evaluated in Section VII-A) while maintaining the same level of expressivity (evaluated in Section VII-B), and to provide clearly defined mechanisms for further extension (illustrated in Section VII-C). Sections VII-B and VII-C compare 27 designs (outlined in Table I) before and after contributions to IRON.

| Name | Description | SLOC Before/After |
|---|---|---|
| **Block Designs** | | |
| Copy ×3 | Pass data unchanged through the NPU using 1) DMAs only, 2) DMAs + kernel, or 3) DMAs + external kernel. | 42/29  48/39 52/42 |
| MTranspose | Transpose a matrix using DMAs. | 36/28 |
| VReduce ×3 | Element-wise vector reduction (add, max, or min). | 50/26  50/26 50/26 |
| VSOp ×2 | Element-wise vector-scalar operation (add or mul). | 49/36  64/55 |
| VVOp ×5 | Vector-vector operations (element-wise add, mul, and mod; vectorized addKern and mulKern). | 56/43  56/43 56/43  94/75 94/76 |
| MSAdd | Scalar addition to sub-matrix in matrix | 62/43 |
| MVAdd | Tiled, row-wise, matrix-vector addition. | 55/44 |
| MVMul | Matrix-vector multiplication. | 105/69 |
| VSoftMax | Softmax implemented using a lookup table approximation, applied to all vector elements. | 79/58 |
| VReLU | Rectified Linear Unit, $ReLU(x) = max(0, x)$, a common DNN activation function. | 76/60 |
| Conv2d ×2 | A 2-dimensional convolution operation; second design fuses kernel with ReLU. | 89/77  86/73 |
| **Advanced Designs** | | |
| GEMM | Matrix-matrix mult. Each matrix is multiplied in blocks with dimensions matching NPU intrinsics, and subtiles distributed to AIEs. | 308/279 |
| BBlock | Bottleneck block used in DNNs such as ResNet [16]. Implemented as a 3 stage pipeline. | 332/245 |
| ResNet Conv2x | Conv2D_X layers of the BBlocks used in ResNet [16] composed of three BBlocks. | 623/414 |
| ColorDetect | Detect two colors in an image, structured as a 3 stage pipeline. | 163/146 |
| EdgeDetect | Detect edges in an image using a 4 stage pipeline. | 198/177 |
| ColorThresh | Tiled data-parallel color threshold detection. | 169/81 |

TABLE I: Twenty-seven designs (21 straight-forward block, 6 optimized advanced) used to evaluate contributions to IRON.

### A. Efficiency

In this section, we decompose the concept of efficiency into two measurements: single lines of code (SLOC) representing brevity and Halstead metrics representing effort. These metrics are imperfect, but we posit the number and variety of evaluated designs imply these results capture representative trends.

*1) Brevity:* SLOC is measured using `pygount` [37] and `radon` [39]. All designs were formatted programmatically using `black` [20]. Figure 4 shows percent decrease between design versions while Table I records the SLOC for each design[2]. Many designs include some level of argument parsing and verification (identical between both versions), and this code slightly dilutes the percentages. Even so, designs written using our contributions to IRON require 25.53% fewer SLOC on average, with an average decrease of 31 SLOC per design.

---
[2]MSAdd SLOC is higher in Table I than in Fig. 2 because the full design includes additional metaprogramming and configuration.

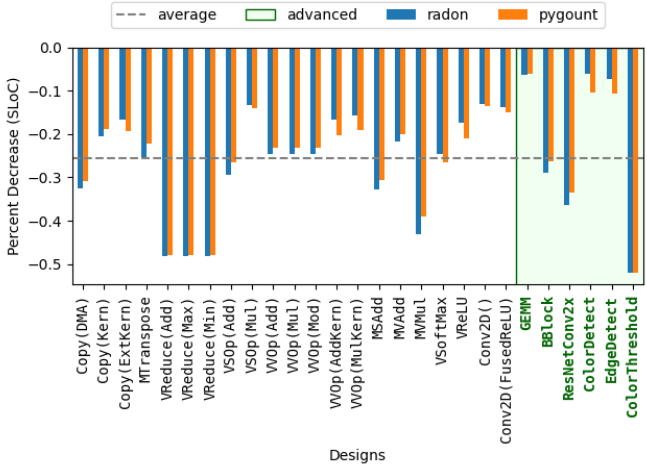

Fig. 4: Average percent decrease of SLOC per design.

*2) Effort:* We use Halstead metrics, a common suite of metrics for evaluating software [4], [7], [12], [14], to analyze the example designs. Halstead metrics use attributes extracted from source code (number of operators, number of operands, and counts of both) to define concepts such as vocabulary and effort [13]. Across all designs for all negative Halstead metrics (calculated using `radon` [39]), the average of the metric is lower for designs written post-contribution to IRON than pre-contribution to IRON. Figure 5 shows the calculations for vocabulary (a measure based on unique operators and operands counted in the source code) and effort (a metric designed to reflect the effort it takes to interact with the program as a writer, maintainer, or reader) [14]. The average reduction for vocabulary is 4.59 (an average percent decrease of 23.32%) while the average reduction in effort is 179.97 (an average percent decrease of 24.82%).

| | L2Mem | SharedMem | Tiling | Broadcast | Split/Join | Window | Skip | Pipeline | Metaprog. | NPUCols |
|---|---|---|---|---|---|---|---|---|---|---|
| GEMM | ✓ | | ✓ | ✓ | ✓ | | | | ✓ | 1-4 |
| BBlock | ✓ | ✓ | | ✓ | ✓ | ✓ | ✓ | ✓ | | 1 |
| ResNet-Conv2x | ✓ | ✓ | | ✓ | ✓ | ✓ | ✓ | ✓ | ✓ | 3 |
| Color-Detect | ✓ | | | ✓ | | | ✓ | ✓ | | 1 |
| Edge-Detect | ✓ | | | ✓ | | ✓ | ✓ | ✓ | | 1 |
| Color-Thresh. | ✓ | | ✓ | | ✓ | | | | ✓ | 1 |

TABLE II: Features of advanced designs.

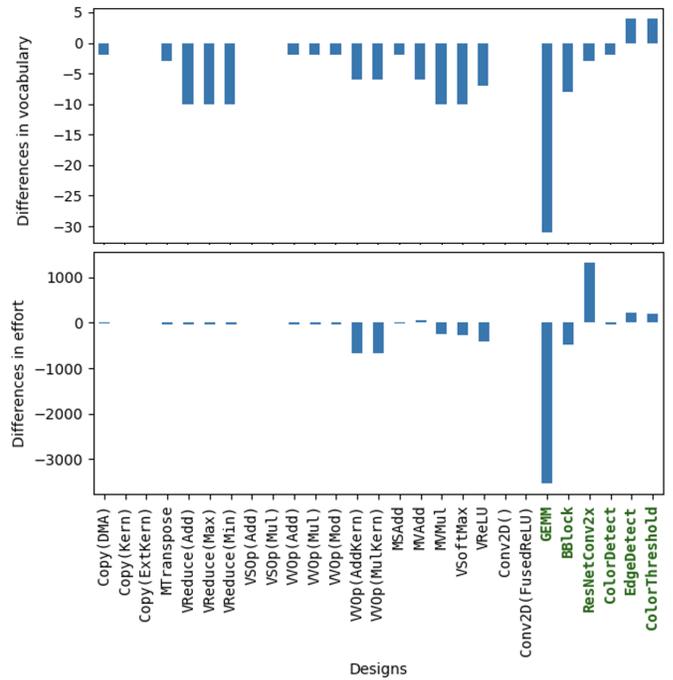

Fig. 5: The difference between Halstead vocabulary (top) and effort (bottom) per design.

### B. Expressivity

This section evaluates whether IRON with contributions is as expressive as IRON before contributions. The example designs, particularly the advanced designs, cover key features and patterns supported by IRON. In Table II, *L2Mem* refers to the use of memory tiles while *SharedMem* refers to the ability of AIEs to access the L1 memory of another AIE (conserving DMA channels). *Tiling* refers to on-the-fly data transformations. *Broadcast*, *Split/Join*, sliding *Window*, and *Skip* connections are features of data movement. *Pipeline* refers to a design which is task-parallel; other designs are data-parallel. *Metaprog* refers to metaprogramming in the design. *NPUCols* reports the number of NPU columns required by the design. All example designs written in IRON post-contribution produce the same (correct) output as the corresponding original designs; this is the first test of correctness. For further analysis, we employ static analysis of MLIR and measure performance data.

To verify both versions of the same design utilize the features of the NPU identically, the generated MLIR from each version is compared. A naive comparison shows generated MLIR is different for all designs. However, most of these differences are in the ordering of declarations (e.g., tiles declared in different order) and so should not affect the functionality of the design. Controlling for the order of declarations, 20 of 27 designs generate identical MLIR. Three designs (`GEMM`, `MVAdd`, `MTranspose`) use different runtime DMA transfer sizes and strides, but the corresponding access patterns expressed are access equivalent and so are representa-tive of functional equivalency. Four designs (`ColorDetect`, `ResNetConv2x`, `BBlock`, `EdgeDetect`) use one or more `ObjectFifo`s with broadcast, and the list of recipients is reordered; reordering does not lead to any differences in functionality.

To corroborate the static analysis, the average latency (100 warm up iterations followed by 1000 iterations) of each design is recorded. The average percentage difference between the versions of each design is approximately 3.36%. 11 of 27 designs using IRON with contributions show (slightly) higher latency than designs using IRON without contributions. The percent difference between the sum of average latencies across designs for each version is less than 0.09%, which is attributed to system noise.

### C. Extensibility

Contributions to IRON include two interfaces designed for extension: the placement interface and `taplib`. We target manual placement and data transformation specification for extension because they can be tedious and error prone tasks, so designers can benefit both from programmability (to algorithmically generate solutions) and the ability to ignore these aspects of a design when desired. In this section, we validate that it is possible to construct extensions using these interfaces.

*1) Defining a `Placer`:* The purpose of this exercise is to show the `Placer` interface a) can be used to generate functional designs, and b) successfully allows a user to create a `Placer` using only knowledge of IRON and NPU architecture. To this end, we create a `SequentialPlacer`, implemented in just 64 SLOC, that uses only constructs from the new IRON API. The `SequentialPlacer` assigns AIEs in a grid-like pattern, and then assigns memory then shim tiles such as to keep `ObjectFifo` endpoints in the same column. This `Placer` is rudimentary; it respects dimensional constraints of the AIE-array but may yield invalid placements due to other resource constraints. Of the 27 example designs, `SequentialPlacer` is used by 24 for full placement; one for partial placement (`BBlock`); and two are fully hand-placed (`ResNetConv2x`, `GEMM`) as the `SequentialPlacer` yielded invalid designs due to over-allocation of resources. This illustrates that the `Placer` API successfully allows new placers to be constructed and applied to IRON designs.

*2) `TensorAccessPattern` Generators:* The true utility of `taplib` primitives is that they may be generated to replace hand-crafted sizes and strides. The intent of a programmer can be difficult to ascertain by viewing the arithmetic used to compute series of sizes and strides. This is problematic for readability and maintainability, and puts the burden for creating such arithmetic on the programmer for every new design. To illustrate how `tap`s and `tases` may by generated for a class of transformations, we implement an example generator, `TensorTiler2D`. `TensorTiler2D` generates `tases` based on tile size, column- or row-wise element access within a tile, groupings of tiles (e.g., access more than one tile in a single `tap`), column- or row-wise access of tiles within the group, group repeats (e.g., access the same pattern more than once within the same `tap`), and group steps (e.g., non-contiguous tile groups of specific step sizes in vertical and horizontal directions). `TensorTiler2D` is implemented in only 277 SLOC and is utilized by five designs (`MSAdd`, `MVAdd`, `MVMul`, `GEMM`, `MTranspose`). We largely attribute the differences in vocabulary and effort in the `GEMM` design (Figure 5) to the `TensorTiler2D`, which greatly reduces the amount of arithmetic in the `Runtime` of this design.

### VIII. RELATED WORK

A class of frameworks targeting NPUs – including AMD Ryzen™ AI Software [5], Intel® OpenVINO™ [34], and Qualcomm® Neural Processing SDK [38] – provide a higher level of abstraction than IRON and focus specifically on machine learning. CUDA [10], OpenCL [33], and HLS [35] are closer to the abstraction level of IRON, but IRON specifically targets accelerators with AIEs. Many works with Python front-ends [8] feature runtime code generation (RCG), often through just-in-time compilation [17], [19], [44]. Others provide domain-specific languages (DSLs) [18] or embedded DSLs (eDSLs) [22], [24], [30]. This work focuses on efficient representation of optimized designs rather than RCG. `Worker` core functions can be considered as written in an eDSL provided by `mlir-python-extras` [26]. Our contributions to IRON instead focus on other aspects of designs, such as data transformations and refinement of constructs representing design structure. [23], [25] proposes a programming model for end-to-end management of AIE architectures, sharing some of our goals (support for performance tuning and metaprogramming) but priorities differ. Our work does not tackle RCG while [23], [25] does not focus on extensible interfaces or measures of designer efficiency. [46] illustrates the complexity of DMA transformations in a scratchpad-based architecture; the tools for expressing DMA transformations provided by `taplib` may be useful for further studies in this area. The use of heat maps and visualization tools for performance engineers is described in [43] and uses similar terminology (access patterns) and techniques (tensor-based heat maps) [43]. However, that work focuses on analyzing the relationship between accesses and movement in cache-based architectures, while access patterns supported by `taplib` in the context of IRON are configurations for data movement. Simulation and modeling described by [43] could be implemented for IRON designs as a future extension.

### IX. CONCLUSION

This work describes, implements, and analyzes contributions to IRON that increase designer efficiency without compromising expressivity, and provide mechanisms for additional extension. The contributions successfully change source code characteristics of 27 example IRON designs, while preserving functionality, through use of a new API with refined programming constructs and extensible interfaces. We observe an average reduction of ∼26% SLOC across example designs due to these changes. While the majority of contributions are specific to IRON, this work provides a data point indicating that fine-tuning the design of programming interfaces for accelerators can have a large effect on characteristics of designs even without changing the abstraction level of the interface.